\documentclass[11pt]{article} 

\usepackage[utf8]{inputenc} 
\usepackage{geometry} 
\usepackage{graphicx} 
\usepackage{fancyhdr} 
\usepackage{sectsty}
\allsectionsfont{\sffamily\mdseries\upshape} 
\usepackage{amsmath}
\usepackage{amssymb}
\usepackage{xcolor}
\usepackage{graphicx}
\usepackage{enumitem}
\usepackage{array}
\usepackage{framed}
\usepackage{bbm}
\usepackage{braket}

\usepackage{comment}

\usepackage{authblk}
\usepackage{changes}

\numberwithin{equation}{section}
\newcommand*{\id}{\mathbbm{1}}

\newcommand{\red}[1]{{\color{red} #1}}
\newcommand{\be}{\begin{equation}}
\newcommand{\ee}{\end{equation}}
\newcommand{\mbf }[1]{{\mathbf{ #1}}}
\newcommand{\hmbf }[1]{\hat{\mathbf{ #1}}}
\title{Teleportation for septuagenarians}
\author{J. E. Avron}
\author{O. Kenneth}
\affil {Department of Physics, Technion, Haifa, Israel}
\date{} 
\begin{document}
\maketitle
\begin{flushright}
{\em To Jurge, Hebert and Tom}
\end{flushright}
\begin{abstract}
An  introduction to the theory of teleportation.
\end{abstract}

\section{Introduction}
Perhaps the first thing one should know about quantum teleportation is that it is a misnomer, at least if by teleportation one means that material objects disappear from one place and reappear elsewhere.  In  quantum teleportation it is the quantum state which is teleported, the quantum data, not its carrier. No qubits or any quantum particle, are ever exchanged between Alice and Bob. Asher Peres, one of the fathers of teleportation, described it as teleportation of the soul without the flesh.    

Quantum teleportation, invented by Bennet, Brassard, Crepau, Jozsa, Peres and Wootters, \cite{teleportation}, is usually   presented as a protocol \cite{nielsen}: Alice and Bob need to do this and that and exchange classical information, so that at the end of the day,  an (unknown) state $\ket{\psi}$ of a  systems of Alice is induced on a system of Bob  without Alice and Bob actually swapping physical systems. 

We present teleporation in a way that splits the mathematics  from the physics. The math part is an identity  which makes use of the (Choi)  isomorphism  between tensor product states  and operators. The identity is a fact in linear algebra and so independent of quantum mechanics.  The physics part is the quantum interpretation of the identity which build on the fact that  in quantum mechanics, a measurement of Alice prepares the quantum state of Bob's system. 

\section{Preliminaries}

\subsection{Baby Choi isomorphism}
Let ${\cal H}$ be a Hilbert space.  We shall denote  $d=\dim{\cal H}$ (possibly infinite).   The (baby) Choi isomorphism \cite{algoet} between (pure) states $\ket{\mbf{C}}$ in ${\cal H}^{\otimes 2}$ and Hilert-Schmidt
operators $\mbf{C}$ on ${\cal H}$: 
\begin{equation}\label{e:B}
\ket{\mbf{C}}=\sum_{j,k=1}^d C_{jk}\ket{j}\otimes \ket{k}=\sum_{k=1}^d \mbf{C}\ket{k}\otimes \ket{k}=\sum_{j=1}^d\ket{j}\otimes {\mbf{C}^t}\ket{j}
\end{equation}
where $\ket{j}\in{\cal H}$ is a basis. 
The  Hilbert-Schmidt condition comes from:
\be\label{e:spd}
\braket{\mbf{C}|\mbf{C}'}=Tr\,  \mbf{C ^\dagger C'} 
\ee
Writing
\begin{equation}\label{e:choy}
\braket{\psi\otimes\varphi |\mbf{C}}=\bra{\psi}\mbf{C}\ket{\varphi^*}
\end{equation}
makes it clear that the isomorphism depends on the anti-isomorphism of ${\cal H}$ and ${\cal H}^*$ determined by picking a basis.

\subsection{Entanglement}
Entanglement of vectors $\ket{\mbf{C}}\in {\cal H}\otimes {\cal H}$, being an intrinsic property of a bi-partite state, does not depend on how Alice and Bob chose bases, is encoded in  unitary invariant properties of the positive part, $|\mbf{C}|=\sqrt{\mbf{C}^\dagger \mbf{C}}$, of the polar decomposition
\be
\mbf{C}=U_\mbf{C}|\mbf{C}|
\ee
The eigenvalues of $|\mbf{C}|$ are known as the Schmidt coefficients,  \cite{nielsen}. Few simple useful facts are: 
\begin{itemize}
\item Maximally entangled states correspond to 
\be\label{e:unitary}
|\mbf{C}|=c\id 
\ee
with $c=1/\sqrt d$ in the case $\mbf{C}$ is normalized.
\item Pure product states correspond to 
\be\label{e:r1}
\text{Rank}\mbf{|C|}= 1
\ee
($\rho_A=\mbf{CC^\dagger }$ is the state of Alice  and  $(\mbf{ C^\dagger C })^t$ of Bob.)

\item For normalized states the von-Neumann entropy 
\be
\dim {\cal H}\ge S(\mbf{|C|})=-Tr \, |\mbf{C}|\log |\mbf {C}|\ge 0
\ee
is  a measure of entanglement.

\end{itemize}
\subsection{Orthogonal bases}
We shall denote   by  $\ket{\mbf{\hat C}_\xi}$ with $\xi\in 1,\dots,d^2$ an orthonormal base in   $ {\cal H}\otimes {\cal H} $:
\begin{equation}
\braket{\mbf{\hat C}_\xi|\mbf{\hat C}_\eta}=
\delta_{\xi,\eta}, \quad  
\sum_\xi \ket{\mbf{\hat C}_\xi}\bra{\mbf{\hat C}_\xi}=\id\otimes \id
\end{equation}
The corresponding relations for $\mbf{\hat  C}_\xi$ are
 \begin{equation}\label{norm-c}
Tr\,\mbf{\hat C}_\xi^\dagger\mbf{\hat C}_\eta=
\delta_{\xi,\eta},\quad \sum_\xi \mbf{\hat C}^\dagger_\xi \mbf{A}\mbf{\hat C}_\xi=\id \  Tr\, \mbf{A}\, 
\end{equation}
Examples are:
\begin{itemize}
\item
A basis of maximally entangled states \cite{teleportation}:
 \begin{equation}\label{e:bme}
\ket{\mbf{\hat C}_{jk}}=\frac 1 {\sqrt d}\sum_a e^{2\pi ijk/d} \ket{a}\otimes \ket{a-j},\quad a,j,k\in\mathbb{Z}_d
\end{equation}
\item A basis of pure product states: 
\begin{equation}\label{e:bpp}
\ket{\mbf{\hat C}_{jk}}=\ket{j}\otimes \ket{k},\quad j,k\in\mathbb{Z}_d
\end{equation}

\end{itemize}

\section{Teleportation identity}

The following identity holds in ${\cal H}^{\otimes 3}={\cal H}_A\otimes {\cal H}_{A'}\otimes {\cal H}_B$:  
\begin{equation}\label{e:ti}
\ket{\psi}_A\otimes \ket{\mbf{C}}_{A'B} = \sum_{\xi} \ket{\mbf{\hat C}_{\xi}}_{AA'}\otimes \mbf{T}_{\xi}\ket{\psi}_B, \quad\mbf{T}_{\xi}= {\mbf{C}^t}\mbf{ \hat C}_{\xi}^\dagger
\end{equation}
The identity ``teleports'' $\ket{\psi}$ from the left (Alice) to the right (Bob).

Proof:  Consider first the finite dimensional case. It is enough to show the identity: 
\begin{equation}\label{e:ti1}
\ket{\psi}\otimes \ket{\id}=  \displaystyle{ \sum_{\xi} \ket{\mbf{\hat C}_{\xi}}\otimes  \mbf{\hat C}_{\xi}^\dagger\ket{\psi}}
\end{equation}
Eq.~(\ref{e:ti}) follows by multiplying both sides on the left by $\id\otimes\id\otimes \mbf{C}^t$ and using Eq.~(\ref{e:B}).

To show  Eq.~(\ref{e:ti1}) note that by linearity it is enough to show it for $\ket{\psi}$ a basis vector $\ket{a}$. Projecting on $\ket{\mbf{\hat C}_\eta}\bra{\mbf{\hat C}_\eta}\otimes \id$ we find
\begin{align*}
 \big(\ket{\mbf{\hat C}_\eta}\bra{\mbf{\hat C}_\eta}\otimes \id \big)\big(\ket{a }\otimes\ket{\id}\big)&=\ket{\mbf{\hat C}_\eta} \otimes\sum_b \braket{\mbf{\hat C}_{\eta}|ab }\ket{b}\\ &=\ket{\mbf{\hat C}_\eta}\otimes\sum_{bc}  \bra{cc} \mbf{\hat C}_\eta^\dagger\otimes\id \ket{ab }\ket{b}\\
&=\ket{\mbf{\hat C}_\eta}\otimes\sum_{b} \bra{b}\mbf{\hat C}_\eta^\dagger \ket{a }\ket{b}
\\
&=\ket{\mbf{\hat C}_\eta}\otimes\mbf{\hat C}_\eta^\dagger \ket{a}.
\end{align*}
The case $d=\infty$  follows from a limit argument. $\hfill \square$
\section{Remote state preparation}

A basic rule of quantum mechanics is that a measurement is a preparation of a quantum state. In particular, a measurement of Alice is a remote preparation of the quantum state  of Bob.

Applied to  the teleportation identity, Eq.~(\ref{e:ti}), this means that if Alice measures her systems to be in the state $\ket{\mbf{\hat C}_\xi}$, she has prepared Bob's state 
\be
\frac{ \mbf{T_\xi}\ket{\psi}}{\mbf{\|T_\xi}\ket{\psi}\|}
\ee
The probability of Alice finding $\xi$, conditioned on the unknown state being $\ket{\psi}$, is :
\be\label{e:prob}
p(\xi|\psi)={\left\| \mbf{T}_\xi\ket{\psi}\right\|^2}
\ee
If Alice communicates to Bob the result of her measurement,  Bob can apply a unitary $\mbf{U}_\xi$ conditioned on $\xi$. The fidelity of teleportation (conditioned on $\ket{\psi}$) weighted by the probability of the event $\xi$  is 
\be\label{e:fid}
 p(\xi|\psi)\left|\frac{ \bra{\psi}\mbf{U_\xi T_\xi}\ket{\psi}}{\mbf{\|T_\xi}\ket{\psi}\|}\right|^2= \left| \bra{\psi}\mbf{U_\xi T_\xi}\ket{\psi}\right|^2\le 
\left| \bra{\psi}|\mbf{ T_\xi}|\ket{\psi}\right|^2
\ee
This shows \cite{banaszek} that the Bob's optimal choice of $\mbf{U}_\xi$ is one that undoes  the unitary in the polar decomposition of  $\mbf{T_\xi}$ (up to an overall phase). The optimal choice induces the state    
\begin{equation}\label{e:bob}
\frac{ |\mbf{T_\xi}|\ket{\psi}}{\mbf{\|T_\xi}\ket{\psi}\|}
\end{equation}
in Bob's system.

\subsection{Ideal teleportation}
In the case that both the initially shared state $\ket{\mbf{C}}$ and the basis   $\ket{\mbf{\hat C}_\xi}$ are maximally entangled  (and normalized), we have,  by Eqs.~(\ref{e:unitary},\ref{e:ti})
\be 
 |\mbf{T}_\xi| \propto \id
\ee
From Eq.~(\ref{e:bob}), Bob recovered  $\ket{\psi}$ with perfect fidelity.  

Note that the probability of Alice finding $\xi$ in this case is by Eqs.~(\ref{e:ti},\ref{e:prob}):
\be\label{e:prob2}
p(\xi|\psi)=\|\mbf{C}^t\mbf{C}_\xi^\dagger\ket{\psi}\|^2=\frac 1 d\|\mbf{C}_\xi^\dagger\ket{\psi}\|^2=\frac 1 {d^2}
\ee
 independent of $\psi$ and $\xi$: Finding $\xi$ gives no information on the state $\ket{\psi}$ \cite{caves}. Had perfect teleportation allowed Alice to learn about the unknown state $\ket{\psi}$, it would allow for non-demolition measurements of unknown states.


 \section{Fidelity of teleportation}
Known quantum state are, in principle,  easy to teleport: Broadcast the preparation protocol. (In the case Alice and Bob are entangled, a single bit suffices, see  \cite{pati}.) Quantum teleportation deals with the case that $\ket{\psi}$ is unknown. In order to evaluate the average fidelity one needs to know the distribution of $\ket{\psi}$. 
In the case $d< \infty$ it is natural to assume that the distribution is uniformly distributed under the unitary group $U(d)$.   We shall denote the corresponding averaging by $\mathbb{E}(\bullet )$.  A standard formula for computing such averages \cite{banaszek}  is given in Appendix \ref{a:average}.

The fidelity of teleportation of a given $\ket{\psi}$ is
\be
F=\sum_{\xi=1}^{d^2}\big(\bra{\psi}|\mbf{T}_\xi|\ket{\psi}\big)^2
\ee
The average fidelity of is, by Eqs.~(\ref{e:fid},\ref{e:average}), \cite{banaszek}
\begin{align}\label{e:af}
\mathbb{E}(F)&=
\frac 1 {d(d+1)}\sum_{\xi=1}^{d^2}\left( Tr\, |\mbf{T_\xi}|^2+\left(Tr\,  \mbf{ |T_\xi|}\right)^2\right)\nonumber \\
&=\frac 1 {d(d+1)} \left( d +\sum_{\xi=1}^{d^2}\left(Tr\,  \mbf{ |T_\xi|}\right)^2\right)
\end{align}
where in the second line we used Eq.~(\ref{norm-c}) to compute the sum over $\xi$ and the normalization of $\ket{\mbf{C}}$, Eq.~(\ref{e:spd}).
Some special cases are:
\begin{itemize}
\item $\mbf{\hat C_\xi}$ are maximally entangled and $\mbf{C}$ any normalized shared state between Alice and Bob: Since 
\be
|\mbf{T_\xi}|=|\mbf{C}^t\mbf{C_\xi^\dagger}|=\frac{|\mbf{C}^t|}{\sqrt d}
\ee
the average fidelity of teleportation 
\be
\mathbb{E}(F)=
\frac 1 {d+1}\left(1+(Tr |\mbf{C}|)^2\right)
\ee
In the case that the shared state $\mbf{C}$ is maximally entangled $Tr |\mbf{C}|=\sqrt d$ and the average fidelity is 1.
In the case that $\mbf{C}$ is a pure product state $Tr |\mbf{C}|=1$ and the average fidelity is $2/(d+1)$.

\item  $\mbf{C}$ a pure product normalized shared state and $\mbf{\hat C}_\xi$ any basis: Since $\mbf{ C}$ is rank one, so is $\mbf{\hat T}_\xi$ and hence 
\be\label{e:tr2}
\sum_\xi \left(Tr\,  \mbf{ |T_\xi|}\right)^2=\sum_\xi Tr\, |\mbf{T_\xi}|^2=d
\ee
It follows from Eq.~(\ref{e:af}) that 
\be\label{e:cl}
\mathbb{E}(F)=\frac 2 {d+1}
\ee

\item
The previous result  has a simple interpretation: Suppose Alice measures $\ket{\psi}_A$  and finds $\ket{j}_A$. The probability for this event is $|\braket{\psi|j}|^2$. She transmits to Bob her result and Bob prepares the state $\ket{j}$.  The fidelity of the preparation with Alice original state is again  $|\braket{\psi|j}|^2$. Hence the fidelity is
\be
F(\ket{\psi})=\sum_j| \braket{\psi|j}|^4=\sum_j \braket{\psi|j}\!\braket{j|\psi}\ \braket{\psi|j}\!\braket{j|\psi}
\ee
and its average, by Eq.~(\ref{e:average}) of the appendix, gives Eq.~(\ref{e:cl}).

\item  $\mbf{C_\xi}$ are pure product states and $\mbf{C}$ any normalized shared state: Since  $\sqrt d \,|\mbf{T}_\xi|$ are rank one projections and Eq.~(\ref{e:tr2}) holds, the average fidelity is
\be
\mathbb{E}(F)=\frac 1 {d(d+1)}\sum_\xi \frac 2 d =\frac 2 {d+1}
\ee
as in the classical setting.

\end{itemize}

 \section{Teleporting with continuous variables}
 When $d=\infty$, there are no maximally entangled states which are normalizable.   In the  case, ${\cal H}=L^2(\mathbb{R}^d)$,  the analog of Bell pairs are  EPR pairs \cite{epr}, represented by distributions. A basis of EPR pairs is parameterized by phase space,  and is given in the coordinate representation, by 
\begin{align}\label{epr-pairs}
\braket{\mathbf{x}_A\otimes \mathbf{x}_B|{\mathbf{q,p}}}
&=\frac {1 }{ (2\pi)^{d/2}}\delta(\mathbf{x}_A-\mathbf{x}_B+\mathbf{q})  e^{i\mathbf{p}\cdot  (\mathbf{x}_A+\mathbf{x}_B)/2}
\end{align}
The extension of teleportation to EPR pairs is due to Vaidman \cite{vaidman}.  The analog of the teleportation identity, Eq.~(\ref{e:ti}), for EPR states is
 \begin{equation}\label{teleportation-c}
\ket{\psi}_A\otimes \ket{{\mbf{q,p}}}_{AB} =
\frac 1 {(2\pi)^d} \int d\mbf{p'}d\mbf{q'} \,e^{i\varphi(\mbf{p,q,p',q'})}\, \ket{\mathbf{q',p'}}_{AA}\otimes  \mbf{T}_{\mbf{q+q',p-p'}}\ket{\psi}_B 
\end{equation}
where $\mbf{T}_{\mbf{q,p}}$ is a phase-space shifts given by:
\begin{equation}
\bra{\mbf{z}}\mbf{T}_{\mbf{q,p}}\ket{\psi}=e^{i\mathbf{p} \cdot \mathbf{z}}\braket{\mbf{z-q}|\psi}\end{equation}
The phase $\varphi$, being an overall phase, is immaterial for teleportation. For the sake of completeness, we list it anyway
\be
2\varphi(\mbf{p,q,p',q'})=\mbf{p \cdot (q'+q)}+\mbf{q' \cdot (p-p')}
\ee 

The case of continuous variables poses several difficulties: First, unlike the finite dimensional case, there is no natural a-priori (finite) measure that represent the notion of a random $\ket{\psi}$ and consequently no natural notion of average fidelity to optimize. Second,   one can not prepare exact EPR states as they are not normalizable.  These difficulties lead to various approximate schemes \cite{braunstein-kimble, furusawa,buzek,paz, enk}.

\appendix
\section{Averaging}\label{a:average}
Known quantum state are, in principle,  easy to teleport: All one needs is broadcast its preparation protocol. Quantum teleportation deals with the case that $\ket{\psi}$ is unknown. In order to evaluate different protocols, it is natural to assume that $\ket{\psi}$ is uniformly distributed under the unitary group $U(d)$.  

To compute averages over $\ket{\psi}$ the following is handy 
\be\label{e:average}
\mathbb{E}(\bra{\psi}\mbf{C}\ket{\psi}\bra{\psi}\mbf{D}\ket{\psi})=\frac 1 {d(d+1)}(Tr\ \mbf{CD}+Tr\,\mbf{C}\  Tr\,\mbf{D})
\ee
To see this note, first, that the average is invariant under unitary transformations of $\mbf{C}$ and $\mbf{D}$ and so the result must be a linear combination of $Tr\ \mbf{CD}$  and ${Tr\,\mbf{C}\  Tr\,\mbf{D}}$.  To see that they come with equal weight write $\ket{\psi}=(\psi_1,\dots,\psi_d)$ and then note that the correlator
\be
\mathbb{E}(\bar\psi_j\psi_k\bar\psi_m\psi_n)\propto \delta_{jk}\delta_{mn}+\delta_{jn}\delta_{mk}
\ee
by Wick theorem, or directly  by symmetry under the exchange $j\leftrightarrow m$ and phase averaging. It follows that 
\be
\mathbb{E}(\bra{\psi}\mbf{C}\ket{\psi}\bra{\psi}\mbf{D}\ket{\psi})\propto \  Tr\ \mbf{CD}+\ Tr\,\mbf{C}\  Tr\,\mbf{D}
\ee
The constant of proportionality  is determined by considering the special case $\mbf{C}=\mbf{D}=\id$. 
\section*{Acknowledgment} The work is supported by ISF.

\end{document}